\documentclass[preprint,11pt,5p]{elsarticle}
\usepackage{amssymb}
\usepackage{graphicx}
\usepackage{amsmath}
\usepackage{amsfonts}
\usepackage[usenames,dvipsnames]{xcolor}
\usepackage{soul}
\biboptions{sort&compress}

\begin{document}
\begin{frontmatter}
\title{A study on dynamics and multiscale complexity of a neuro system}
\author[A2]{Sanjay K. Palit}
\author[A3]{Sayan Mukherjee \corref{cor1}}

\cortext[cor1]{Corresponding author, Email: msayan80@gmail.com}

\address[A2]{Basic Sciences and Humanities Department, Calcutta Institute of Engineering and Management, Kolkata, India}
\address[A3]{Department of Mathematics, Sivanath Sastri College, Kolkata, India}

\begin{abstract}
We explore the chaotic dynamics and complexity of a neuro-system with respect to variable synaptic weights in both noise free and noisy conditions. The chaotic dynamics of the system is investigated by bifurcation analysis and $0-1$ test. A multiscale complexity of the system is proposed based on the notion of recurrence plot density entropy. Numerical results support the proposed analysis. Impact of music on the aforesaid neuro-system has also been studied. The analysis shows that inclusion of white noise even with a minimal strength makes the neuro dynamics more complex, where as music signal keeps the dynamics almost similar to that of the original system. This is properly interpreted by the proposed multiscale complexity measure.
\end{abstract}
\begin{keyword}
Neuro dynamics \sep Power noise \sep $0-1$ test \sep Recurrence plot \sep Music signal
\end{keyword}
\end{frontmatter}
\section{Introduction}
\label{intro}
An artificial neural network (ANN) is a mathematical model analogical with the biological structure of a neuron, which consists of a cellular body with a dense centroid of activity called the nucleus, entering nerves that receive signals from other neurons called dendrites and the departing nerves that carry signals away from the neurons called axons \cite{mod1}. It is represented by a directed graph composed of neurons as the nodes, nerves or synapses as the edges and an algorithm describing the conduction of impulses through the network. The extent to which the input of neuron $i$ is driven by the output of the $j$ neuron is characterized by its output and the synaptic weight $w_{ij}$. Positive value of the synaptic weight $w_{ij}$ indicates that the output of the neuron $j$ excites the neuron $i$, while the negative value indicates the output of the neuron $j$ inhibits the neuron $i$. If the output of the neuron $j$ has no influence on the neuron $i$, then the synaptic weight $w_{ij}$ equals zero \cite{mod2}. \par
The human neural system is very much complex and its complex dynamic evolutions \cite{modd2} that lead to chaos have already been observed experimentally. Most of the theoretical models of neural systems exhibit stable and cyclic behaviors, yet there also exists some models that illustrate the existence of chaos in neural networks. These models rely on complex architectures or complex equations for both neuron and synaptic dynamics to display chaos. Sometimes the quantities which exhibit chaotic evolutions in these models have no direct physiological interpretations. In \cite{mod3}, chaos in neural networks appears for the evolution of the sum of the absolute values of the synaptic weights of a network. A wide range of studies on small networks has been made by different investigators. Glass et al. discussed a transition from steady state through limit cycle to chaos for networks of six or more neurons \cite{mod4}. In \cite{mod5}, it has been demonstrated that the onset of chaos in an eight neuron system and numerically track down the transition from steady state through limit cycles to chaos. In \cite{mod6}, different dynamical regimes has been reported, particularly the evidence of possibility of chaotic regimes in individual neuron output activity. They have shown the transition of the system from a stable to a chaotic regime as synaptic weight increases. In \cite{mod7}, authors have shown a detailed numerical simulations on how the stability of the system passes from stable state to chaotic state and also discussed some biological implications. They have also made an attempt to find the parameters on which the stability of the system depends most sensitively. \par
During the past few decades, complexity analysis of deterministic and stochastic systems has become an integral part of nonlinear analysis. In all kinds of real world phenomena, some sort of uncertainty is always being there. Obviously, for a stochastic phenomenon it is more than a deterministic phenomenon. This actually means that as the system becomes more and more random, the amount of uncertainty gradually increases. This is measured by entropy, first introduced by C.E. Shannon \cite{en1}. More is the entropy value, more uncertainty is there in the corresponding phenomenon. The term complexity is used in this context. In general complexity is positively correlated with entropy. Since the inception of Shannon entropy, several entropy measures have been developed \cite{en2,en4,en5,en6,en7} and used widely in diverse domains of research \cite{sb1,sb2,sm1}.

After the introduction of the recurrence plots (RP) \cite{rp1,rp2,rp3}, few other measures of complexity \cite{rp4,rp41,rp42,rp43} have been introduced. All of these measures were found to be more effective even than the Lyapunov exponent for the determination of the divergence behavior of dynamical systems. In RP, various structures provide different information regarding the nature of phase space. Diagonal lines describe parallel movements, while trapping situation/ laminar states are described by vertical/horizontal lines. Presence of only diagonal lines with equal/unequal time span indicates periodicity/quasi-periodicity of the phase space. Chaotic regime can be understood from rectangular like structure consists of diagonal lines with some isolated points and vertical lines. All of these basic features of the phase space can be characterized by Recurrence period density (RPD). The idea of RPD is based on recurrent time between the recurrent points. Shannon entropy of recurrence times is called Recurrence plot density entropy (RPDE) \cite{RPD}, which is found to be very effective to calculate the degree of complexity of the phase space. However, a multiscale approach \cite{rp5,rp6,rp7,rp8,rp9,rp10} of the RPDE has not been explored so far, which is expected to reflect the dynamical characteristics of complex systems more accurately.\par
In this article, the dynamics of the three neuron systems \cite{mod7} has been further investigated in noise free, noise induced and music perturbed condition to look after the dynamical changes of the system. The dynamics is quantified by single and two parameter bifurcation diagrams followed by $0-1$ test \cite{ch1,ch2,ch3,ch4}. The $0-1$ method measures underlying chaotic structure of the system from one of its solution component (time series), whatever the system is deterministic or noise-induced \cite{ch5}. Chaos in noise-induced system has already been established in \cite{ch6,ch7,ch8,ch81}. The $0-1$ test is based on mean square displacement (MSD), measured from the diffusive and non diffusive part of a time series and can be applied for deterministic as well as stochastic dynamics \cite{ch2}. The MSD is found to be a bounded function of time for regular dynamics, while it scales linearly with time for chaotic states. The asymptotic growth ($K_c$) of MSD serves as a measure to quantify the dynamics of a system or a time series. For chaotic and regular dynamics, $K_c$ comes close to $1$ and $0$ respectively. The main advantage of $0-1$ test is that it does not require any phase space reconstruction that depends on finding proper time-delay and embedding dimension of the time series. For this reason, the test is found to be suitable for the analysis of discrete maps, ordinary differential equations, delay differential equations, partial differential equations and real world time series. The test can be applied even for time series contaminated with noise \cite{ch5}. Thus, $0-1$ test stands as one of the most promising alternative measures of standard Lyapunov exponent methods to the analysis of discretely sampled data. Moreover, it does not involve any kind of preprocessing of the data and needs only a minimal computational effort independent of the dimension of the underlying dynamical system under investigation. $0-1$ test has found its applications in a wide range of fields that includes but not limited to the studies of dissipative, Hamiltonian dynamical systems, multi-agent systems,  various engineering, electronics, finance and economics, geophysical applications, hydrology, epidemiology and traffic dynamics \cite{ch3,ch4}. The test is even applicable to non-smooth processes, to systems with fractional derivatives and delays, and to non-chaotic strange attractors, where standard methods of computing Lyapunov exponents cannot be applied \cite{ch3,ch4}. The results show a strong correlation between $K_c$ and bifurcation analysis. In order to know the long term characteristics of the systems, multiscale RPDE is proposed, which strongly correlates with $K_c$. Finally, this multiscale RPDE is used to explore the changes in complexity of the neuro systems in noise and music perturbed condition.
\section{Dynamics of three neurons}
\label{sec2}
\subsection{Three dimensional neural network model}
\label{sec11}
\begin{figure}[h]
\begin{center}
  \includegraphics[width=3.2in,height=2.0in,trim=0.0in 0in 0in 0in]{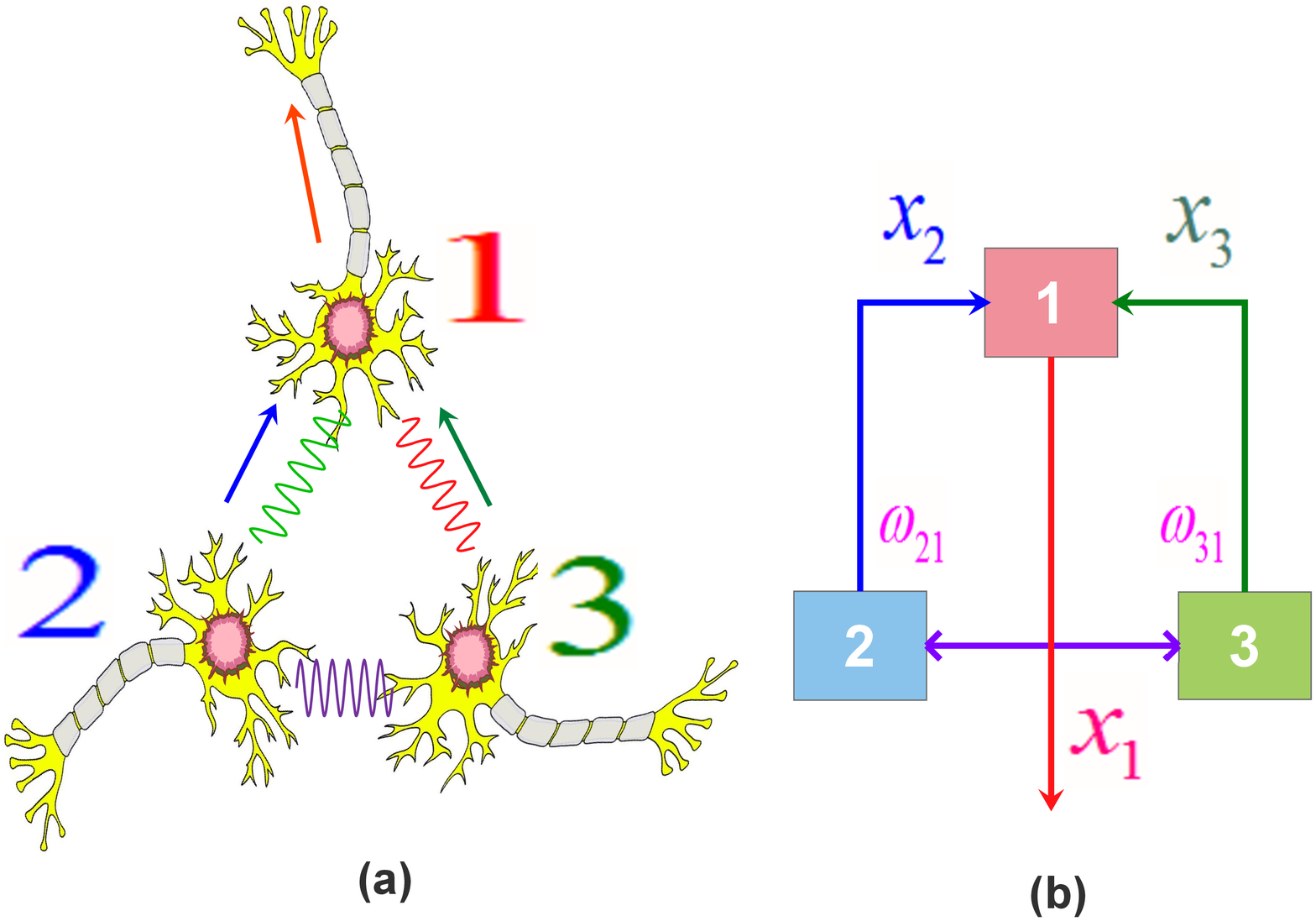} %[width=4.8in,height=3.0in,trim=0.0in 0in 0in 0in]
\end{center}
\caption{(a) represents three connected neurons-$1,2,3$. Arrow indicates the direction of the output generated by the neuron. (b) represents schematic diagram of a three neuron network. $x_1,x_2,x_3$ indicates output of the respective excited neurons 1, 2 and 3.}
\label{fig:Fig1}       % Give a unique label
\end{figure}

Let $x_1,x_2, x_3$ respectively denotes the output activity of the three neurons $1, 2, 3$. The weights of the synaptic connections from neuron $2$ to $1$ and neuron $3$ to $1$ are denoted by $w_{21}$ and $w_{31}$ respectively. The corresponding schematic diagram is given in Fig.\ref{fig:Fig1}. 
With each neuron, there is associated a non-negative bounded (bounded by $0,1$) sigmoidal response function given by $f_i(s)=(1+e^{-\beta_i(s-\theta_i)})^{-1}, i=1,2,3$, where $\beta_i,\theta_i$ respectively denotes the slope and the threshold of the response function for the neuron $i$. The equations of control for this sequence of events with the response function $f_i$ is described by Das et. al \cite{mod7}. The corresponding noise induced system is given by
\begin{align}\label{eq:eq1}
\frac{dx_1}{dt}&=f_1(w_{21}x_2+w_{31}x_3)-\alpha_1 x_1+K\phi(\xi(t)),\\\nonumber
\frac{dx_2}{dt}&=f_2(x_1)-\alpha_2x_2,\\\nonumber
\frac{dx_3}{dt}&=f_3(x_1)-\alpha_3x_3,
\end{align}
where where $\alpha_1,\alpha_2,\alpha_3$ are the respective decay rates, assumed to be constant. $K$ is the noise strength and $\phi(\xi)$ is the Gaussian white noise.  For the entire simulation, we choose $\alpha_1=0.52,\alpha_2=0.42,$ and $\alpha_3=0.1$.
\subsection{Bifurcation and $0-1$ test}
In this section, we investigate the dynamics of (\ref{eq:eq1}) with individual as well as combined effect of $w_{21}$ and $w_{31}$. The investigation is done in both noisy and noise free conditions.
\label{sec12}
In this section, we investigate the dynamics of (\ref{eq:eq1}) with individual as well as combined effect of $w_{21}$ and $w_{31}$. The investigation is done in both noisy and noise free conditions.
\subsubsection{Individual effect of $w_{21}$ and $w_{31}$} 
\begin{figure}[h]
\begin{center}
  \includegraphics[width=3.4in,height=3.2in,trim=0.0in 0in 0in 0in]{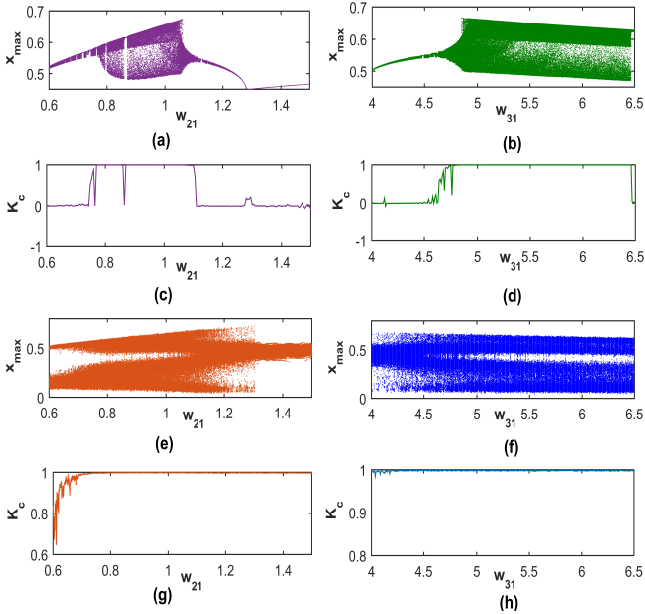} %[width=4.8in,height=3.0in,trim=0.0in 0in 0in 0in]
\end{center}
\caption{(a), (b) respectively represents the bifurcation diagrams of the Neuro system (1) in noise free condition ($K=0$) for varying synaptic weights $w_{21} \in [0.4,1.5], w_{31} \in [4,6.5]$. (c), (d) represents $w_{21}$ and $w_{31}$ vs. $K_c$ graphs with fixed $w_{31}=5.2$ and $w_{21}=1$ respectively. (e), (f) respectively represents the bifurcation diagrams for the same range of parameter values of $w_{21},w_{31}$ in noisy condition with noise strength $K=0.05$. (g), (h) respectively represents $w_{21}$ and $w_{31}$ vs. $K_c$ graphs with fixed $w_{31}=5.2$ and $w_{21}=1$ respectively in noisy condition.}
\label{fig:Fig2}       % Give a unique label
\end{figure}

We first investigate the bifurcation scenario of (\ref{eq:eq1}) with the changes of $w_{21}, w_{31}$. Fig.\ref{fig:Fig2}a, b shows the corresponding bifurcation diagrams for $K=0$ with $w_{21} \in [0.4,1.5],w_{31}=5.2$ and $w_{31} \in [4,6.5], w_{21}=1$ respectively. Fig.\ref{fig:Fig2}a shows single/double and multiple periods for $w_{21} \leq 0.75, w_{21}>0.75$ respectively. However, the multi-periodicity is lost for $w_{21}>1.1$. It indicates that region of multiple and single/quasi-periodic behavior can be observed in $[0.75,1.1]$ and $[0.6,1.5]\setminus[0.75,1.1]$ respectively. 
On the other hand, the system (\ref{eq:eq1}) shows periodic/quasi-periodic behavior for $w_{31}<4.75$ but becomes multi-periodic with the increase of $w_{31}$ as evident from Fig.\ref{fig:Fig2}b. Similar analysis has been done with $K=0.05$. The corresponding bifurcation diagrams are given by Fig.\ref{fig:Fig2}e and f respectively. It is seen from Fig.\ref{fig:Fig2}e and f that the system always possesses multiple periods for $w_{21}\in [0.6,1.5]$,with $w_{31}=5.2$ and $w_{31}\in [4,6.5]$ with $w_{21}=1$. Since bifurcation analysis is done only for finding-‘period route to chaos’, the above analysis can only indicate that the dynamics of the noise-induced system (\ref{eq:eq1}) has a higher tendency of producing chaotic like structures for a wider range of parameter values than the same in noise-free condition. \vskip 3pt
To investigate regular (periodic/quasi-periodic) and chaotic behavior of the system (\ref{eq:eq1}), we have used $0-1$ test method. In this method, only one solution components, say $x(j), j=1,2,..,N$ is translated by
\begin{equation}
p_c(n)=\sum_{j=1}^{n}x(j)\cos(jc),\quad q_c(n)=\sum_{j=1}^{n}x(j)\sin(jc),
\end{equation}
where $c \in (0,\pi)$ and $n=1,2,..,N$.\\
The diffusive and non-diffusive behavior of $p_c$ and $q_c$ is then investigated by measuring mean square displacement (MSD) $M_c$ \cite{ch1,ch2} given by,
\begin{equation}
M_c=\lim_{N\to\infty} \frac{1}{N}\sum_{j=1}^{N}[p_c(j+n)-p_c(j)]^2+[q_c(j+n)-q_c(j)]^2,
\end{equation}
where $n<<N$. The limiting value of $M_c$ is assured only for $n \leq n_{cut}$, where $n_{cut} <<N$. For the practical purpose, $n_{cut}=\frac{N}{10}$ reveals good result \cite{ch1,ch2}.  
In order to investigate the behavior of $M_c$, the asymptotic growth $K_c$ of $M_C$ is calculated by
\begin{equation} \label{eq:asymp}
K_c=\lim_{n\to\infty} \frac{\log M_c(n)}{\log n}.
\end{equation}
The value of $K_c$ close to $1$ and $0$ indicates chaotic and regular dynamics respectively \cite{ch1,ch2}. \vskip 3pt
For numerical simulation, we have considered $x$-components of (\ref{eq:eq1}). Fig.\ref{fig:Fig2}c, d represents the fluctuation of $K_c$ with $K=0$ under the variables $w_{21}$ (keeping fixed $w_{31}=5.2$) and $w_{31}$ (keeping fixed $w_{21}=1$) respectively.
It can be observed from the Fig.\ref{fig:Fig2}c that $K_c$ is close to $0$ and $1$ for $w_{21}\in[0.6, 0.75)\cup (1.1,1.5]$ and $w_{21}\in[0.77, 0.86]\cup [0.87,1.1]$ respectively. On the other hand, it can be observed from Fig.\ref{fig:Fig2}d that $K_c$ is close to $0$ for $w_{31}<4.63$ and $w_{31}\in [4.77,6.46]$, while $K_c$ comes close to $1$ for $w_{31}\in (6.46, 6.5]$. Thus, the fluctuations of $K_c$ can quantify the chaotic as well as the non-chaotic regime of (\ref{eq:eq1}) for the variable synaptic weights $w_{21},w_{31}$ respectively. Similar investigation is done with $K=0.05$. The corresponding fluctuations are shown in Fig.\ref{fig:Fig2}g and h respectively. From the figures, it can be observed that the respective values of $K_c$ are close to $1$ and hence indicates chaos for $w_{21}\in [0.62, 1.5],w_{31}=5.2$ and $w_{31}\in [4, 6.5],w_{21}=1$. Therefore, inclusion of white noise with a small strength can enhance the chaos in a certain range of parameter space. As chaotic dynamics is a signature of complex phenomenon in a system, it assures greater paradigm of complex dynamics exists in noise-induced system compared to the same in noise-free condition.

%Therefore, inclusion of white noise with a very small strength makes the Neuro-system more complex. In fact, the noise perturbed system becomes chaotic almost everywhere in the aforesaid ranges even where it was periodic/quasi-periodic in noise free condition.

\subsubsection{Combined effect of $w_{21}$ and $w_{31}$ } 
\begin{figure}[h]
\begin{center}
  \includegraphics[width=3.5in,height=2.2in,trim=0.0in 0in 0in 0in]{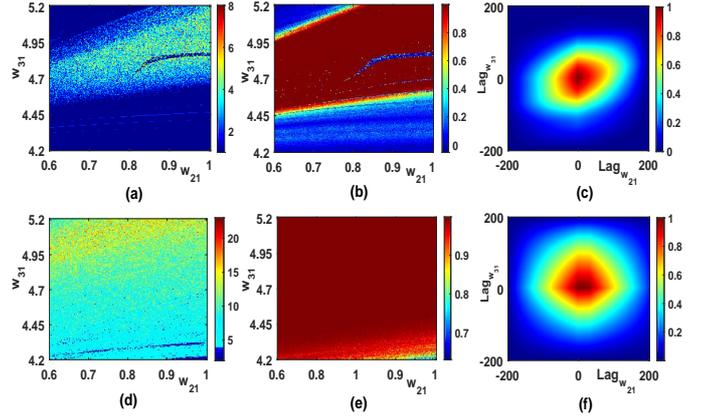} %[width=4.8in,height=3.0in,trim=0.0in 0in 0in 0in]
\end{center}
\caption{(a), (d) respectively represents the 2D bifurcation diagrams and contour diagram representing $K_c$ values for the Neuro-system (1) in noise free condition ($K=0$) with varying synaptic weights $w_{21} \in [0.4,1.5], w_{31} \in [4,6.5]$. (b), (e) respectively represents the same in noisy condition with noise strength $K=0.05$. (c),(f) represent the 2D cross correlation diagram of (a),(b) and (d), (e) respectively. The associate color bars indicate values of the cross-correlation.}
\label{fig:Fig3}       % Give a unique label
\end{figure}

We first investigate two parameter bifurcation of the system (\ref{eq:eq1}) with $K=0, 0.05$. The corresponding diagrams are shown in Fig.\ref{fig:Fig3}a,d respectively. 
It can be observed from Fig.\ref{fig:Fig3}a that the system exhibits multiple periods (3 or more) in the region $[0.6, 1]\times [4.55, 5.2]-[0.6, 0.75]\times [4.97, 5.2]-[0.82, 1]\times [4.75, 4.8]$. On the other hand, Fig.\ref{fig:Fig3}d shows that the multi-periodicity occurs almost everywhere in the region $[0.6, 1]\times [4.2, 5.2]$. So, the inclusion of the white noise with $K=0.05$ increases the number of periods of the Neuro-system than the same with $K=0$. 
The chaotic and non-chaotic region is then classified by using $0-1$ test under the variable parameters $w_{21},w_{31}$. 
The contour diagram in Fig.\ref{fig:Fig3}b and e represent the variation of $K_c$ values with respect to $w_{21},w_{31}$ respectively. The $K_c$ values in Fig.\ref{fig:Fig3}b indicates that the system is chaotic in the range $[0.6, 1]\times [4.55, 5.2]-[0.6, 0.75]\times [4.97, 5.2]-[0.82, 1]\times [4.75, 4.8]$ in noise free condition. However in noise induced condition, chaotic dynamics is observed almost everywhere in the region $[0.6, 1]\times [4.25, 5.2]$  as evident from Fig.\ref{fig:Fig3}e. Therefore, the white noise even with a minimal strength has a strong influence on the system and it makes the system chaotic irrespective of the synaptic weights $w_{21},w_{31}$. To check whether or not the bifurcation analysis and $0-1$ test lead to same type of conclusion regarding the dynamical pattern of the Neuro-system, 2D correlation analysis is further performed with respect to different lags of $w_{21},w_{31}$. This are given by Fig.\ref{fig:Fig3}c and f for noise free and noise induced condition respectively. Both the 2D correlation diagrams show strong correlation between the two parameters bifurcation and $0-1$ test for the Neuro-system. Thus, 2D correlation analysis confirms that as the number of periods increases, the neuro system loses its stability and leads to chaos in both noise free and noisy conditions.\par
\begin{figure}[h]
\begin{center}
  \includegraphics[width=3.3in,height=2.2in,trim=0.0in 0in 0in 0in]{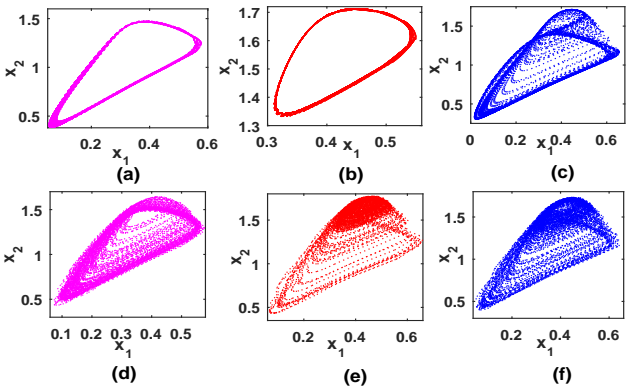} %[width=4.8in,height=3.0in,trim=0.0in 0in 0in 0in]
\end{center}
\caption{(a), (b), (c) respectively represents the 2D projection of the phase space of the neuro system of for different combination of synaptic weights $w_{21}=0.75,w_{31}=5.2;w_{21}=1,w_{31}=4.5;w_{21}=1,w_{31}=5.2$ in noise free condition ($K=0$). (d), (e), (f) respectively represent the similar diagrams in noisy condition ($K=0.05$).}
\label{fig:Fig4}       % Give a unique label
\end{figure}

We next investigate the asymptotic dynamics of (\ref{eq:eq1}) directly from its phase space in noise free ($K=0$) and noise induced condition ($k=0.05$). Some of the prominent cases in form of 2D projection of the phase diagrams are presented in Fig.\ref{fig:Fig4}a-f. Fig.\ref{fig:Fig4}a-f again confirm that the inclusion of white noise with a very small strength makes the dynamics chaotic even where it was periodic/ quasi- periodic in noise free condition.
\section{Multiscale complexity in noise-induced neuro-system} 
\label{sec3}
\subsection{Recurrence plot and multiscale normalized Recurrence period density entropy}
\label{sec31}
Recurrence in a $n$-dimensional phase space $X=\{(\vec{x}_{i}):\vec{x}_{i}\in\Re^{n}, i=1,2,...,N\}$, indicates the closeness of its points. Two points $x_{i},x_{j}\in X, i=1,2,...,N$ are considered close i.e. recurrent if $\|\vec{x}_{i}-\vec{x}_{j}\|<\epsilon$. The corresponding recurrent matrix is defined as 
\begin{equation}
R_{i,j}=\Theta (\epsilon-\|\vec{x}_{i}-\vec{x}_{j}\|), i=1,2,...,N,
\end{equation}
where $\Theta$ is the Heaviside function, $\| .\|$ is the Euclidean norm of the phase space, and $\epsilon$ denotes the radius of the neighborhood. The symbols `1' (black dots) and `0' (white dots) are used to represent the recurrent and non-recurrent points respectively.
Recurrent time denoted by $T_{k}$ is computed as the number of non-recurrent points or white lines between two recurrent points $x_{i},x_{j}$ in the RP $R_{i,j}$. Formally, recurrent time for a pair of recurrent points $x_{i},x_{j}\in R_{i,j}$ is defined as $T_{k}=(i-j)$. Thus, $T_{1}$ corresponds to the least recurrent time,$T_{2}$ corresponds to the next and so on. A series of recurrent time interval $n(T_{k})$ for all points in $R_{i,j}$ is obtained as the number of occurrence of $T_{k}$. RPD denoted by $P(T_{k})$ is defined as the probability of $n(T_{k})$ among the sample space $\{n(T_{k})\}$. This is given by $(8)$.
\begin{equation}
P(T_{k})=\frac{n(T_{k})}{\sum_{k=1}^{T_{max}}n(T_{k})},
\end{equation}
where $T_{max}=max \{T_{k}\}$.
RPD can quantify the complexity of the phase space. However, it can not measure the order of complexity. This is done by a RPD based entropy called Normalized Recurrence period density entropy (NRPDE).
Recurrence periodic entropy (RPDE) of the reconstructed phase space, where the points are independently identically distributed is defined by utilizing the concept of Shannon entropy \cite{en1}. Thus, RPDE is given by
\begin{equation}
H=-\sum_{k=1}^{T_{max}}P(T_{k}) \log P(T_{k}).
\end{equation}
Since $T_{max}$ varies with sampling time, a normalization of RPDE is necessary. The normalized RPDE (NRPDE) is defined as
\begin{equation}
H_{norm}=-(log T_{max})^{-1}\sum_{k=1}^{T_{max}}P(T_{k}) \log P(T_{k}).
\end{equation}
Here $\log(T_{max})$ is equal to the entropy of a purely random variable, given by
$$\log (T_{max})=-\sum_{k=1}^{T_{max}}P(T_k)\log P(T_k),$$where $P(T_k)\sim\frac{1}{T_{max}}$.\vskip 3pt
To measure the order of complexity more accurately, MNRPDE is defined by utilizing the MAV multiscaling technique \cite{mav} on the NRPDE $H_{norm}$ as follows:\vskip 3pt
For the time series $x$ (defined as above), the multiscale time series, denoted by $\{z_j^{(s)}\}_{j=1}^{N-s+1}$ is defined as
\begin{equation}
z_j^{(s)}=\frac{1}{s}\sum_{i=j}^{j+s-1}x_i
\end{equation}
For each scale $s$, we can define the multiscale NRPDE $H_{norm}^{(s)}$ by Eq.(10). The mean of $\{H_{norm}^{(s)}\}_{s=1}^{s_{0}}$ is then defined by 
\begin{eqnarray}\label{eq4}
 <H_{norm}>=\frac{1}{s_0}\sum_{s=1}^{s_{0}}H_{norm}^{(s)}, 
\end{eqnarray}
where $<.>$ represents statistical average. \vskip 3pt
In the following section, we verify the effectiveness of $<H_{norm}>$ by measuring the dynamical complexity of (\ref{eq:eq1}).
\subsection{Complexity in neuro system under variable synaptic weights}
\begin{figure}[h]
\begin{center}
  \includegraphics[width=3.5in,height=2.5in,trim=0.0in 0in 0in 0in]{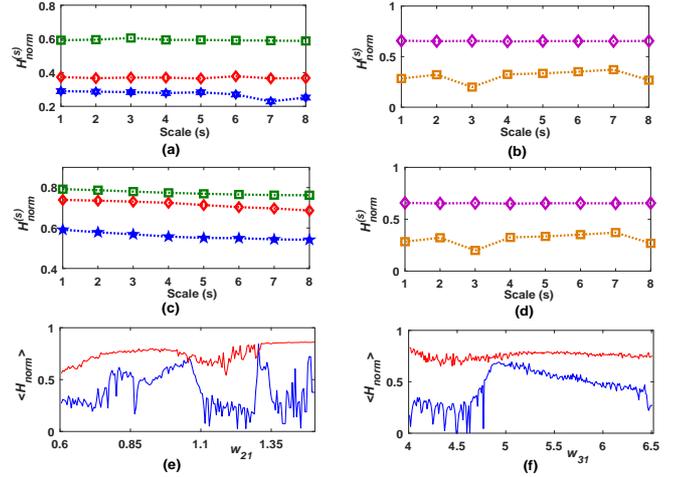} %[width=4.8in,height=3.0in,trim=0.0in 0in 0in 0in]
\end{center}
\caption{(a), (b) respectively represents the graph of MNRPDE $-$ $H_{norm}^{(s)}$ for some fixed value of the synaptic weights $(w_{21},w_{31})=(0.6,5.2),(1,5.2),(1.1,5.2)$ and $(w_{21},w_{31})=(1,4.1),(1,5.1)$ in noise free condition ($K=0$). (c), (d) respectively represents the similar graphs in noisy condition ($K=0.05$).(e) represents $<H_{norm}>$ for varying $w_{21}\in [0.6,1.5]$ with a fixed $w_{31}=5.2$ in noise free (blue line) and noise induced (red line) conditions. (f) represents the same plot for varying $w_{31}\in [4,6.5]$ with a fixed $w_{21}=1$. RP is constructed from the attractor reconstructed from $x_1$ component of the solution vector with embedding dimension 3 and time-delay 10.}
\label{fig:Fig5}       % Give a unique label
\end{figure}

\begin{figure}[h]
\begin{center}
  \includegraphics[width=3.8in,height=2.6in,trim=0.0in 0in 0in 0in]{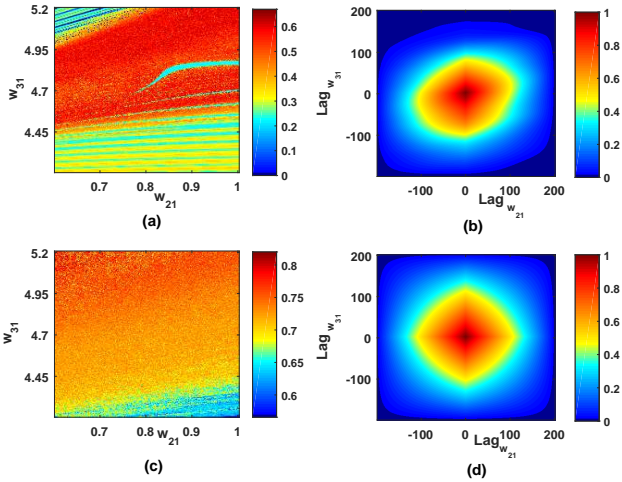} %[width=4.8in,height=3.0in,trim=0.0in 0in 0in 0in]
\end{center}
\caption{(a), (c) respectively represents the contour plots of $-$ $<H_{norm}>$ for varying synaptic weights $w_{21} \in [0.6,1], w_{31} \in [4.2,5.2]$ in noise free ($K=0$) and noise induced ($K=0.05$) conditions. (b), (d) respectively represents the 2D cross-correlation of the two parameter $<H_{norm}>$ plot with two parameter $0-1$ test plot in noise free and noise induced conditions. Color bars indicate values of the cross-correlation.}
\label{fig:Fig6}       % Give a unique label
\end{figure}

To measure the dynamical complexity, we have first investigated the multi-scaling behavior of (\ref{eq:eq1}) using $H_{norm}^{(s)}$ with the scale $s=1,2,..,8$. This is given by Fig.\ref{fig:Fig5}. 
Fig.\ref{fig:Fig5}a, c show the fluctuations of $H_{norm}^{(s)}$ for fixed $(w_{21},w_{31})=(0.6,5.2),(1,5.2),(1.1,5.2)$ in both noise free and noisy conditions respectively, while Fig.\ref{fig:Fig5}b, d represent the similar graphs for fixed $(w_{21},w_{31})=(1,4.1),(1,5.1)$. From the Fig.\ref{fig:Fig5}a-d it can be observed that $H_{norm}^{(s)}$ gives different values for different scales. Thus, the mean value of $H_{norm}^{(s)}$ is expected to reflect the degree of complexity of the neuro system properly. Fig.\ref{fig:Fig5}e and f respectively shows the variation of $<H_{norm}>$ over variable $w_{21}, w_{31}$ in both noise free and noisy conditions. It can be seen from the figures that the degree of complexity increases for the neuro system in noisy condition with respect to both the parameters. This correlates with the earlier results of bifurcation analysis and $0-1$ test. We next investigated the behavior of $<H_{norm}>$ under the combined effect of $(w_{21},w_{31}) \in [0.6,1]\times [4.2,5.2]$ in both noise free and noisy conditions. The corresponding matrix plots are given in Fig.\ref{fig:Fig6}a and c respectively. Comparing these plots with the same in Fig.\ref{fig:Fig3}b and e, it can be observed that both of $<H_{norm}>$ and $K_c$ plots are almost similar for same set of parameter values of $w_{21},w_{31}$ in noise free and noisy conditions. The correlation between them has also been investigated. Fig.\ref{fig:Fig6}b and d represents respective 2D correlation contour, which establishes almost correlated patterns between $<H_{norm}>$ and $K_c$.
\section{Application on the music perturbed neuro system}
\begin{figure}[h]
\begin{center}
  \includegraphics[width=3.3in,height=2.2in,trim=0.0in 0in 0in 0in]{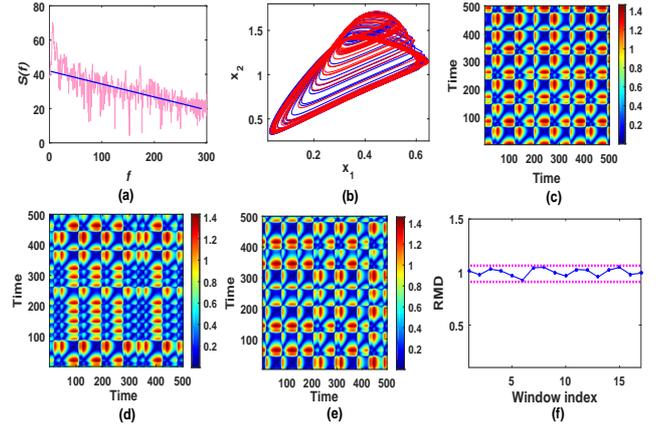} %[width=4.8in,height=3.0in,trim=0.0in 0in 0in 0in]
\end{center}
\caption{(a) represents the graph of power spectral density of the music signal with respect to variable frequencies. (b) represents the joint attractors of the neuro system (1) (blue) and the corresponding music perturbed system (11) (red). (c), (d), (e) represent three samples of sub distance matrix plots. The associate color bars represents values of $d_{ij}$ between the points $(x_i,y_i)$. (f) represents the graph of $RMD_{i} (R_i)$ for different window index $i$. The distance matrix $(d_{ij})_{N\times N}$ thus obtained is then subdivided into $m=[\frac{N}{500}]$ sub matrices, each of size $500$.}
\label{fig:Fig7}       % Give a unique label
\end{figure}

In this section, we investigate chaotic dynamics and complexity of the system (\ref{eq:eq1}) under an effect of music signal. For the numerical experiment, we have considered an instrumental music signal $Mu(t)$ with power $S(f)=\frac{1}{f^\alpha}$. Fig.\ref{fig:Fig7}a shows corresponding $f$ vs. $S(f)$ graph. From the figure, it can be observed that the slope $\alpha$ of the line representing the mean trend of $S(f)$ is approximately $2$. So $\alpha=2$. The music perturbed system of (\ref{eq:eq1}) is given by
\begin{align}\label{eq:music_neuro}
\frac{dx_1}{dt}&=f_1(w_{21}x_2+w_{31}x_3)-\alpha_1 x_1 +K_1 Mu(t),\\\nonumber
\frac{dx_2}{dt}&=f_2x_1-\alpha_2x_2,\\\nonumber
\frac{dx_3}{dt}&=f_3x_1-\alpha_3x_3.
\end{align}
where $K_1$ denotes the strength of the music.\par
Fig.\ref{fig:Fig7}b  shows the attractors of the neuro system (\ref{eq:eq1}) with $K=0$ (blue) and the music perturbed neuro system (\ref{eq:music_neuro}) (red) with $w_{21}=1,~w_{31}=5.2$ and $K_1=0.05$. It is observed that the dynamical pattern of both the attractors are almost similar. To quantify this, we measure distance  $d_{ij}=\|x_i-y_j\|$ for different windows $W_s$ with $w_{21}=1,~w_{31}=5.2$, where $x_i, y_i (i,j=1,2,...,N)$ respectively denotes the $i,j^{th}$ point on the attractors of neuro systems (1) ($K=0$) and (11). The windows are defined by $W_s=\{(d_{i,j})_{M_s \times M_s}: M_s \leq N\}$. Fig.\ref{fig:Fig7}c, d, e show three such window matrix plots as sample illustrations. It can be observed that $d_{ij} \in [0,1.4]$ for all $i,j$ in each case. As $d_{ij}$ indicates dispersion between the trajectories of (1) ($K=0$) and (11), its corresponding windows reflect changes between the respective attractors. We define a ratio $R=\frac{\bar{W}_s}{\bar{W}_{s-1}}$, where $\bar{W}_s=\frac{1}{N^2} \sum_{i=1}^{M_s} \sum_{j=1}^{M_s} d_{ij}$ ($d_{ij} \in W_s$ and $M_s \leq N)$. We call $R$ by ratio of mean distance (RMD). Naturally, $R \approx 1$ only when two consecutive windows possess the same mean. It implies that average distance between the trajectories of the respective systems (1) ($K=0$) and (11) does not vary over time. Fig.\ref{fig:Fig7}f shows the values of $R$ (RMD) over $s=1,2,..,8$. It is observed that the $R \approx 1$ for all $s$ and hence proves that system (1) ($K=0$) and (11) have the similar trajectory movements with $w_{21}=1,w_{31}=5.2$.

\begin{figure}[h]
\begin{center}
  \includegraphics[width=3.3in,height=1.4in,trim=0.0in 0in 0in 0in]{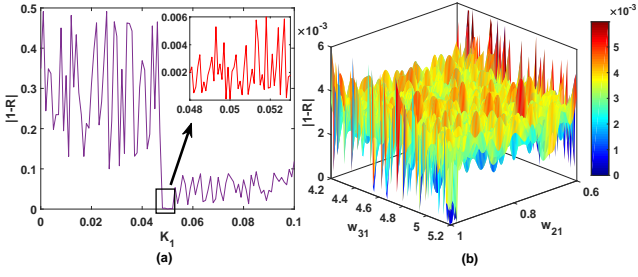} %[width=4.8in,height=3.0in,trim=0.0in 0in 0in 0in]
\end{center}
\caption{(a) represents $|1-R|$ vs. $K_1 \in [0.0.1]$ graph for the system (11) with $w_{21}=1,~w_{31}=5.2$. (b) represent surface of $|1-R|$ over the region $(w_{21},w_{31}) \in [0.6,1] \times [4.2,5.2]$ with fixed $K_1=0.05$ for the same system.}
\label{fig:Fig7A}       % Give a unique label
\end{figure}

Keeping fixed $w_{21}=1,~w_{31}=5.2$, we further investigated the same dispersion between the trajectories over $K_1 \in [0,0.1]$. The corresponding $|1-R|$ vs. $K_1$ graph is shown in Fig.\ref{fig:Fig7A}a. From the figure, it can be observed that values of $|1-R| \approx 0$ for $K_1 \in [0.048,0.053]$. It can verified that $|1-R|=0$ for $K_1=0.0495,0.05$. It implies $R=1$, i.e; almost similar phase spaces can be obtained for the systems (1) and (11) at $K_1=0.0495,0.05$ with $w_{21}=1,~w_{31}=5.2$. Further, oscillation of $|1-R|$ is calculated over the region $(w_{21},w_{31}) \in [0.6,1] \times [4.2,5.2]$. The corresponding surface is given in Fig.\ref{fig:Fig7A}b. From Fig.\ref{fig:Fig7A}b, it can be investigated that values of $|1-R| \leq 0.006$ for all $(w_{21},w_{31}) \in [0.6,1] \times [4.2,5.2]$ with fixed $K_1=0.05$. It assures that the system (1) ($K=0$) and (11) possess almost similar phase spaces with the changes in $(w_{21},w_{31}) \in [0.6,1] \times [4.2,5.2]$ (for fixed $K_1=0.05$).

\begin{figure}[h]
\begin{center}
  \includegraphics[width=3.4in,height=1.0in,trim=0.0in 0in 0in 0in]{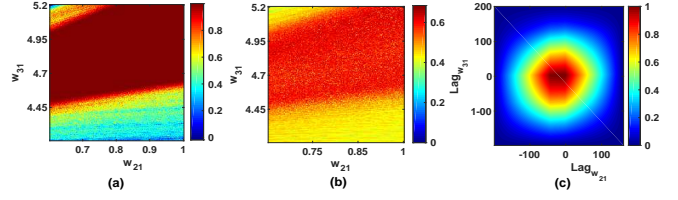} %[width=4.8in,height=3.0in,trim=0.0in 0in 0in 0in]
\end{center}
\caption{(a) $K_c$ vs. $(w_{21},w_{31}) \in [0.6,1] \times [4.2,5.2]$ graph with $K_1=0.5$ for the system (11). (b) represent fluctuation of $<H_{norm}>$ under the variation of $(w_{21},w_{31}) \in [0.6,1] \times [4.2,5.2]$ with $K_1=0.5$ for the same system. In (c), correlation between the $K_c$ (shown in Fig.\ref{fig:Fig8}a) and $<H_{norm}>$ (shown in Fig.\ref{fig:Fig8}b) with $Lag_{w_{21}} \in [-200,200],~Lag_{w_{31}}  \in [-200,200]$. For (a)-(c), the respective color bars indicates values of the $K_c$, $<H_{norm}>$ and correlation.}
\label{fig:Fig8}       % Give a unique label
\end{figure}

In the next, we thus investigated chaotic dynamics and complexity in the dynamics of (\ref{eq:music_neuro}) under the variation of $(w_{21},w_{31})$ with fixed $K_1=0.05$. The chaotic dynamics is characterized using $0-1$ test method. To do this, we have calculated fluctuation in $K_c$ with $(w_{21},w_{31}) \in [0.6,1] \times [4.2,5.2]$ ( for fixed $K_1=0.05$). The corresponding matrix plot is shown in Fig.\ref{fig:Fig8}a. The dark color in Fig.\ref{fig:Fig8}a, corresponds $K_c \approx 1$. It verifies existence of chaotic dynamics in (11). Further, complexity is measured by calculating $<H_{norm}>$ over same $(w_{21},w_{31}) \in [0.6,1] \times [4.2,5.2]$ with fixed $K_1=0.05$. Fig.\ref{fig:Fig8}b shows corresponding matrix plot. From Fig.\ref{fig:Fig8}a and b, similar patterns can be observed between the respective fluctuation in $K_c$ and $<H_{norm}>$. To confirm the similarity, we have done a 2D cross-correlation analysis. The cross-correlation contour is given in Fig.\ref{fig:Fig8}c. From Fig.\ref{fig:Fig8}c, it can be investigated that cross-correlation is almost equal to $1$ at $(Lag_{w_{21}},Lag_{w_{31}})=(0,0)$. It assures strong correlation between $K_c$ and $<H_{norm}>$ under the variation of $(w_{21},w_{31}) \in [0.6,1] \times [4.2,5.2]$ (for fixed $K_1=0.05$). However, respective dynamical changes as well as complexity between the noise free system (1) ($K=0$) and noise induced system (1) ($K=0.05$), and also between (1) ($K=0$) and (11) cannot be classified from this study.  

\begin{figure}[h]
\begin{center}
  \includegraphics[width=3.3in,height=1.4in,trim=0.0in 0in 0in 0in]{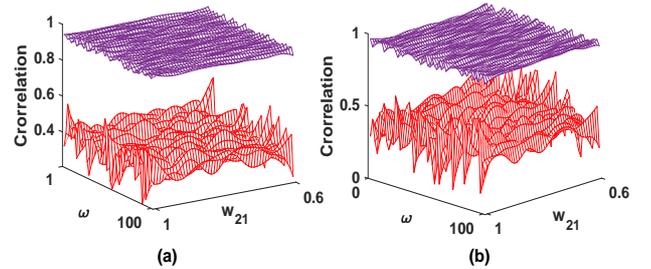} %[width=4.8in,height=3.0in,trim=0.0in 0in 0in 0in]
\end{center}
\caption{(a) represents correlation values for $K_c$ in $CaseI$ (in red color) and $CaseII$ (in violet color) at each $w_{31}=\omega \in [4.2,5.2]$ under the variation $w_{21} \in [0.6,1]$. (b) represents correlation values for $<H_{norm}>$ in the aforesaid cases at each $w_{31}=\omega \in [4.2,5.2]$ over $w_{21} \in [0.6,1]$. To calculate the CRs, we have considered 100 fixed values of $\omega \in [4.2,5.2]$. }
\label{fig:Fig9}       % Give a unique label
\end{figure}

To classify the changes, we have considered two hypotheses:\\
$$H_0/\mathbb{A}: \mathbb{A}/CaseI=\mathbb{A}/CaseII$$
$$H_1/\mathbb{A}: \mathbb{A}/CaseI\not =\mathbb{A}/CaseII$$
$$H_0/\mathbb{B}: \mathbb{B}/CaseI=\mathbb{B}/CaseII$$
$$H_1/\mathbb{B}: \mathbb{B}/CaseI\not =\mathbb{B}/CaseII,$$
where $\mathbb{A},~\mathbb{B}$ denotes the event for $K_c$ and $<H_{norm}>$ respectively. $\mathbb{A}/CaseI$ and $\mathbb{A}/CaseII$ stands for the respective standard deviations of the samples $CaseI$, $CaseII$. Here, $CaseI$ indicates correlation between the system (1) with $K=0$ and  the same with $K=0.05$. Similarly, $CaseII$ indicates the same between the systems (1) with $K=0$ and (11). In order to find the correlation, we calculate cross-correlation (CR) at zero lag for each $w_{31}=\omega \in [4.2,5.2]$ under the variation of $w_{21} \in [0.6,1]$. Fig.\ref{fig:Fig9}a and b shows surfaces of CR for $K_c$ and $<H_{norm}>$ respectively with $w_{31}=\omega \in [4.2,5.2],~w_{21} \in [0.6,1]$. It can be observed from Fig.\ref{fig:Fig9} that, CR$\geq 0.95$ for $CaseII$. On the other hand, the same CR$\leq 0.56$ for $CaseI$. It indicates weak and strong correlation for the $CaseI$ and $CaseII$ respectively.  

Further, two sample $t$-test confirms that both $H_1/\mathbb{A}$ and $H_1/\mathbb{B}$ are true with $p (< 0.00001)$ significance level. It confirms stronger correlation in $CaseII$ than the same in $CaseI$ with $w_{31}=\omega \in [4.2,5.2],~w_{21} \in [0.6,1]$. So, hypothesis testing shows that dynamical as well as complexity patterns of the noise free neuro system (1) (with $K=0$) are highly correlated with the music perturbed system (11) compared to the noise induced system (1) (with $K \not =0$).   
\section{Conclusions}
In this article, the dynamics and complexity of a neuro system both have been studied under noise free, noisy and music perturbed conditions. To investigate complex dynamics, bifurcation analysis is done only for noise free and noise induced systems. The results indicate that larger number of multi-periods exist in the noise induced system compared to the same in noise free condition, whatever may be the variation in both synaptic weights. Further, $0-1$ test shows chaotic paradigm in the noise induced system is greater than the same in noise free condition under the same synaptic variation.
The proposed multiscale entropy $<H_{norm}>$ shows a strong correlation with $K_c$ in both noise free and noisy conditions. So, $<H_{norm}>$ can reflect the complex nature of neuro dynamics properly. The neuro system is then perturbed with an instrumental music. It has been observed that the dynamics of the music perturbed system has a close similarity with the original neuro system. Since music has a soothing effect on human feeling and mood, the inclusion of music signal with the neuro system keeps the dynamics almost unchanged. To investigate this, distances between every pair of points on the attractors of the respective original and music perturbed neuro system are computed. Based on these distance window based ratio RMD is then defined which clearly establishes the similarity between the dynamics. Fluctuation of both $K_c$ and $<H_{norm}>$ are finally investigated for a certain range of parameter values $w_{21}$ and $w_{31}$. Both of them reflect the actual changes in the dynamics of the noise free, noise induced and music perturbed neuro systems. In fact, it assures similarity between the dynamics of the original (noise free) and music perturbed neuro systems, while they show dissimilarity in the dynamics of the original and noise induced neuro systems. Finally two samples $t$-test hypothesis confirms that almost similar dynamics can be obtained in the case of music perturbed dynamics compared to the noisy neuro system.
Thus, our newly proposed measure $<H_{norm}>$ can properly interpret the complexity of the neuro dynamics in noise free, noisy and music perturbed conditions. Since the values of $<H_{norm}>$ of the original and music perturbed neuro systems are found to be almost same for variable synaptic weights $w_{21}, w_{31}$ and an optimal music strength $K_1=0.05$, $<H_{norm}>$ also reflects the soothing effect of music on the neuro system. The present study also reveals that the soothing effect of music will be destroyed if $K_1<0.05$ as $|1-R|$ highly deviates from $0$ in this range. However, $|1-R|$ shows a mixed trend for $K_1>0.05$ and thus it needs further investigation on how the neuro system reacts on music perturbation in this case. This is definitely a future scope of the present research. 

\end{document}